\documentclass{IOS-Book-Article}

\usepackage{mathptmx}
\usepackage{graphicx}
\usepackage{amssymb,amsmath,caption}
\usepackage{subcaption}
\usepackage{amsmath}
\usepackage{mathtools}
\usepackage{tablefootnote}
\usepackage{footnote}
\usepackage{booktabs}
\usepackage{array} 
\usepackage{url}
\usepackage{color}
\usepackage[section]{placeins}
\usepackage{fancyhdr}

\def\hb{\hbox to 10.7 cm{}}

\begin{document}

\pagestyle{headings}
\def\thepage{}

\newcommand{\sap}{Al$_{2}$O$_{3}$}
\newcommand{\lithium}{$^6$Li }

\begin{frontmatter}              % The preamble begins here.

%\pretitle{Pretitle}
\title{In-Beam Background Suppression Shield}

\markboth{}{December 2015\hb}
%\subtitle{Subtitle}

\author[A]{\fnms{V. Santoro},} %
\author[A,B]{X.~X.~Cai,}
\author[A,C]{D.~D.~DiJulio,}
\author[A]{S.~Ansell }
and
\author[A,D]{P.~M.~Bentley}

\runningauthor{B.P. Manager et al.}
\address[A]{European Spallation Source ERIC, SE-221 00 Lund, Sweden}
\address[B]{DTU Nutech, Technical University of Denmark, Roskilde, Denmark}
\address[C] {Department of Physics, Lund University, SE-221 00 Lund, Sweden} 
\address[D] {Department of Physics and Astronomy, Uppsala University, 751 05 Uppsala, Sweden} 

\begin{abstract}
The long (3ms) proton pulse of the European Spallation Source (ESS) gives rise to unique and potentially high backgrounds for the instrument suite. 
In such a source an instrument capabilities will be limited by it's Signal to Noise (S/N) ratio.
The instruments with a direct view of the moderator, which do not use a bender to help mitigate the fast neutron background, are the most challenging.
For these beam lines we propose the innovative shielding of placing blocks of material directly into the guide system, which allow a minimum attenuation of the cold and
 thermal fluxes relative to the background suppression. This shielding configuration has been worked into a beam line model using Geant4.
 We study particularly the advantages of single crystal sapphire and silicon blocks .   

\end{abstract}

\begin{keyword}
background, sapphire, filter, neutrons, spallation source
\end{keyword}
\end{frontmatter}
\markboth{December 2015\hb}{December 2015\hb}

\section{Introduction}
\begin{figure}[h] \centering
   	\includegraphics[width=12 cm]{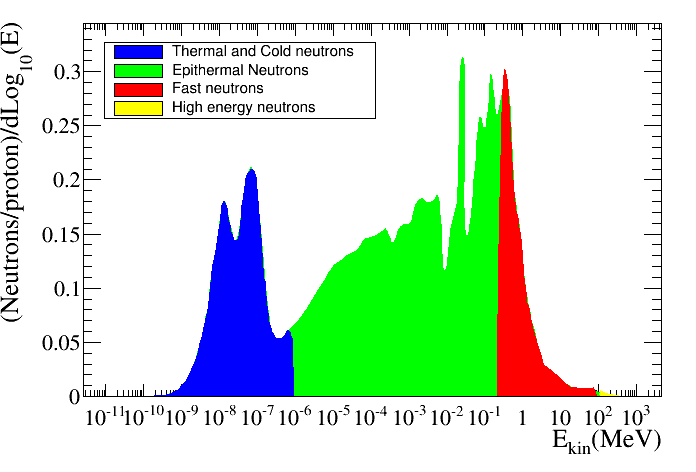}
	\caption[ESS source]{ESS TDR~\cite{r1} moderator spectrum. The blue band corresponds to neutrons of energy below 1 eV, the green one to neutrons from 1 eV to 100 keV, the red one to 100 keV to 10 MeV, while the yellow band to neutrons with energy greater than 10 MeV.}
	\label{fig:modspectrum}
 \end{figure}

\indent The European Spallation Source ESS~\cite{r1} is being constructed in Lund, Sweden, and is planned to be the world brightest pulsed neutron source. The facility uses a 2 GeV proton beam hitting a tungsten target to produce neutrons (predominately evaporation neutrons at about 2~MeV and spallation product that are considered background). The neutrons are then moderated in moderators consisting of both liquid hydrogen and water compartments. Surrounding the moderators are multiple beamports that use neutron guides to transport cold ( \textless~20~meV [millielectron volt]) and thermal neutrons to the sample position of an instrument
via reflection, over a distance of several tens of meters for a short instrument and a hundreds meters for a long instrument.
As well as the desired neutron flux, many higher energy neutrons ($>$ eV, up to 2 GeV) and other particles also scatter down the beamlines, via scattering with the guides,
shielding, choppers and other thermal neutron optical components.
This gives rise to potentially high backgrounds for the instruments, limiting their capabilities by their Signal to Noise (S/N) ratio.  One of the methods of reducing the fast particle background is a ``T0 chopper", which is a large rotating mass that passes through and obscures the neutron beam during the accelerator proton pulse.  T0 choppers are expensive, and mechanical points of failure that require maintenance, so we are exploring steady-state alternatives.\\  
\indent One possibility is to place blocks of material directly into the guide system,  which allows for  a minimum attenuation of the cold and
 thermal fluxes relative to the background suppression. These devices are called filters since they have a wavelength dependent cross-section in such a way that is low for thermal but strongly increases at epithermal and high energies.
 The choice of material and its dimension are critical parameters affecting the performance of the filter. In this work, we studied sapphire (\sap) and Silicon (Si), two materials that are usually employed at neutron facilities since they are  characterised by having a total cross section capable of attenuating the unwanted epithermal neutrons and gamma radiation in the incident spectrum~\cite{sapphirefiltersel1, sapphirefiltersel2}.
In this study, these materials have been tested using the ESS spectrum at the beam port entrance, which is roughly located 2~m after the moderator. 
The target and the moderator design used to calculate this spectrum corresponds to the ESS TDR design ~\cite{r1,dougKon}.
 The spectrum is shown in Fig.~\ref{fig:modspectrum}. As can be seen by the different energy bands, the spectrum has a large energy range that goes from cold and thermal (blue band) to high energy neutrons (yellow band).
 The red band (from 100 keV to 100 MeV ) corresponds to the fast neutron component while the green bands corresponds to neutrons from 1 eV to 100 keV, which are called epithermal neutrons and this is the region where nuclear resonances occur in nuclei.\\  
 \indent For the purpose of our study we classify as ``background" all the neutrons with energy greater than 1~eV and as ``Signal" all the neutrons with energy less than 1~eV, since the instruments at ESS will not use any neutrons with energy $>$ 0.5 eV.
In this work we simulated with Geant4~\cite{geant4package} different thicknesses and different geometrical configurations for \sap and Silicon filter in order to optimise the S/N ratio.

\section{Geant4 and NCrystal package} 	

Geant4 is a Monte-Carlo toolkit for the simulation of the passage of
particles through matter. It is one of the codes used at ESS for shielding~\cite{ourpaper1,ourpaper2,ourpaper3,ourpaper4,ourpaper5}
and detector simulations. The toolkit is developed by a worldwide
collaboration of physicists, implemented in C++ and has an open source
license.  The code is used across a number of different scientific
fields, including high-energy physics, accelerator physics and medical
and space science.  The physics processes offered cover a
comprehensive range, from meV and extending to TeV energies, and
include electromagnetic, hadronic and optical processes with a large
set of long-lived particles, materials and elements. 

The composition of a material is flexible since it is possible to specify the atomic and isotopic content of a given volume.
However, in Geant4 the effects on the interaction cross-section for thermal neutrons due to inter-atomic bindings is only present for commonly 
used materials in nuclear engineering.  For this reason, at ESS a package called NCrystal has been developed in order to simulate slow neutron transport in polycrystals and
imperfect single crystals. The crystal structure can be supplied by the user or by  the nxs library~\cite{r3,r4,r5}.
The cross section is then calculated and applied for coherent and incoherent neutron scattering.
This package is a key element of this work since the wanted filter properties (attenuation of epithermal or higher energy neutrons with almost no effect on thermal and cold neutrons) are described by this tool included in the
ESS detector framework~\cite{griff}. In this current work, NCrystal is combined with the Neutron HP models for neutrons with an energy range from 4 eV to 20 MeV, which are based on evaluated neutron data libraries, and also with the Bertini \cite{Bertini63,Bertini69} model for neutrons with energies greater than 20MeV.

 \begin{figure}[h]
 \centering
   	\includegraphics[width=12 cm, height=6cm]{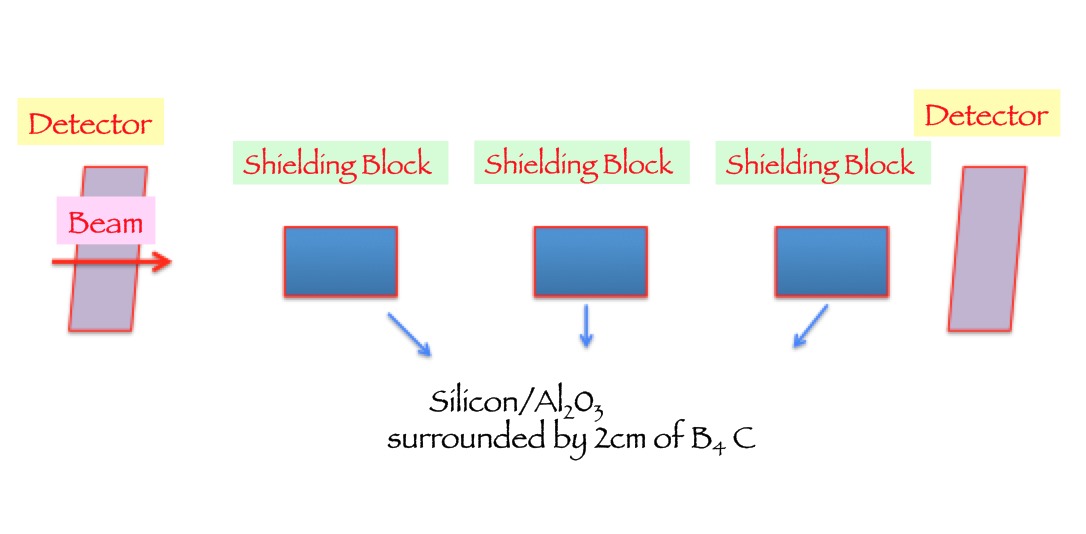}
	\caption[ESS source]{Schematic view of the simulation setup.}
	\label{fig:geant4sim}
 \end{figure}

\begin {table}[h]
\centering
\caption {Attenuation Coefficients and Signal to Noise ratios for different configurations of Silicon filters.}
\label{tab:sil} 
\begin{tabular}{ c c c c}
\hline\hline
Configuration & Attenuation Coefficient  & Attenuation Coefficient & S/N ratio \\  
 &  for High Energy Neutrons & for Low Energy Neutrons & \\  
No Si & - & - & 0.3 \\
2 cm & 2.5 & 1.7 & 0.44  \\  
5 cm & 2.5 & 1.7 & 0.44  \\  
10 cm & 3.9 & 1.9 & 0.62  \\  
2 $\times$ 10 cm & 9 & 2.4 & 1.2 \\  
3 $\times$ 10 cm & 18.& 3 & 1.9  \\  
3 $\times$ 15 cm & 40 & 3.9 & 3.1  \\  
3 $\times$ 18~cm & 54 & 4.6 & 3.6  \\  
3 $\times$ 24~cm & 82 & 6 & 4  \\  
3 $\times$ 27~cm & 99 & 7.3 & 4.1  \\  
3 $\times$ 30~cm & 117 & 8.6 & 4  \\  
3 $\times$ 40~cm & 179 & 14 & 3.74  \\  
3 $\times$ 50~cm & 251 & 23 & 3.2  \\  
\hline
\hline
\end{tabular}
\footnotetext{} 
\end{table}

\begin {table}[h!]
\centering
\caption {Attenuation Coefficients and Signal to Noise ratios for different configurations of Sapphire filters. }
\label{tab:sap} 
\begin{tabular}{ c c c c}
\hline\hline
Configuration & Attenuation Coefficient  & Attenuation Coefficient & S/N ratio \\  
 &  for High Energy Neutrons & for Low Energy Neutrons & \\  
No \sap & - & - & 0.3 \\
2 cm & 2.99 & 1.79 & 0.51  \\  
5 cm & 8.28 & 2.24 & 1.13  \\  
10 cm & 40.90& 3.18 & 3.91  \\  
2 $\times$ 10 cm & 479 & 6.02 & 24.24 \\  
3 $\times$ 10 cm & 1632 & 10.9 & 45.54  \\  
3 $\times$ 15 cm & 4206 & 25.5 & 50.03 \\  
3 $\times$ 18~cm & 6160 & 41.81 & 44.59 \\  
3 $\times$ 30~cm & 12483 & 278 & 13.52\\  
\hline
\hline
\end{tabular}
\footnotetext{} 
\end{table}

\begin{figure}
\centering
      \begin{subfigure}[t]{0.35\textwidth}
         \centering
               \includegraphics[width=6.8cm,height=5.8cm]{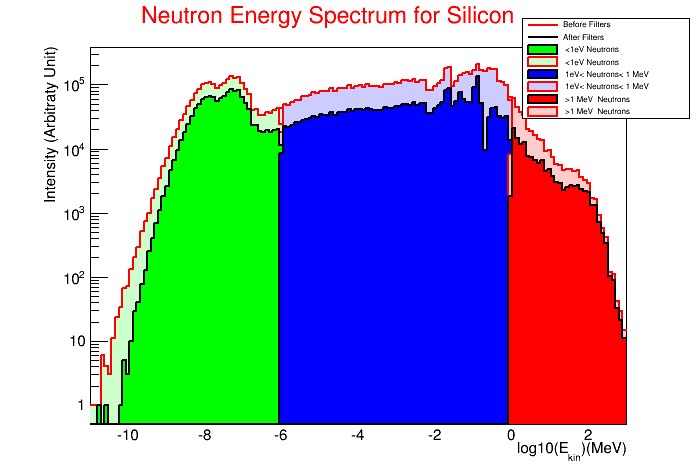}
        \caption{Neutron spectrum before and after the filter for 1 block of 2 cm. }
        \label{fig:sia}
    \end{subfigure}
    \hspace{2.5cm}
       \begin{subfigure}[t]{0.35\textwidth}
        \includegraphics[width=6.8cm,height=5.8cm]{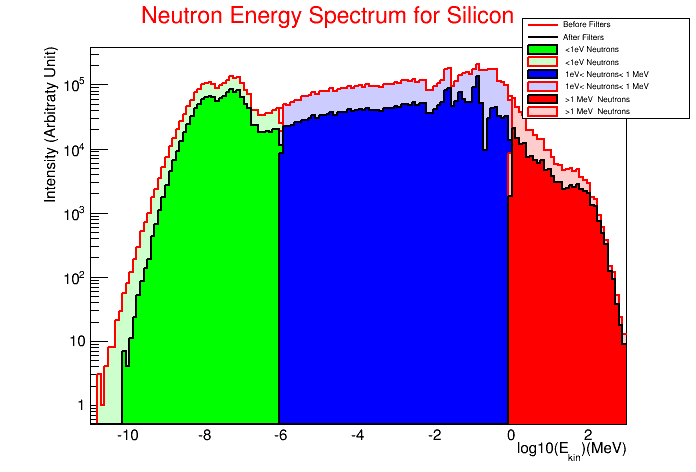}
        \caption{Neutron spectrum before and after the filter for 1 block of 5 cm.}
         \label{fig:sib}
       \end{subfigure}
         \begin{subfigure}[t]{0.35\textwidth}
        \includegraphics[width=6.8cm,height=5.8cm]{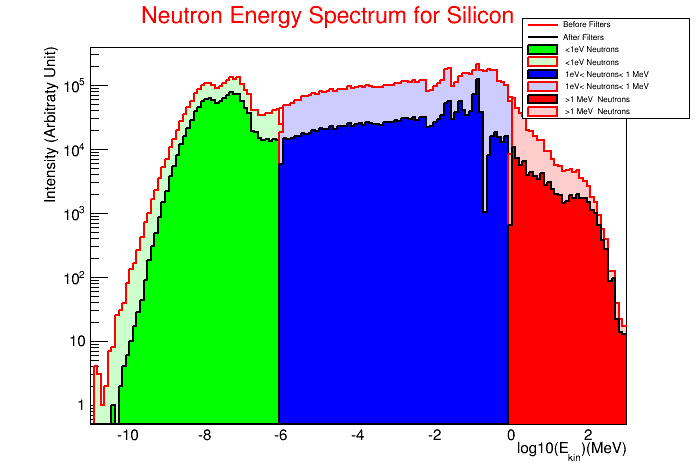}
        \caption{Neutron spectrum before and after the filter for 1 block of 10 cm.}
        \label{fig:sic}
        \end{subfigure}
        \hspace{2.5cm}
       \begin{subfigure}[t]{0.35\textwidth}
        \includegraphics[width=6.8cm,height=5.8cm]{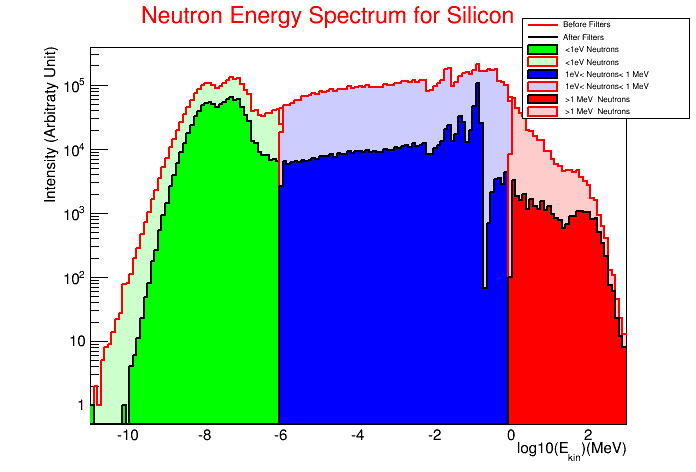}
        \caption{Neutron spectrum before and after the filter for 2 blocks of 10 cm.}
         \label{fig:sid}
    \end{subfigure}
    \begin{subfigure}[t]{0.35\textwidth}
        \includegraphics[width=6.8cm,height=5.8cm]{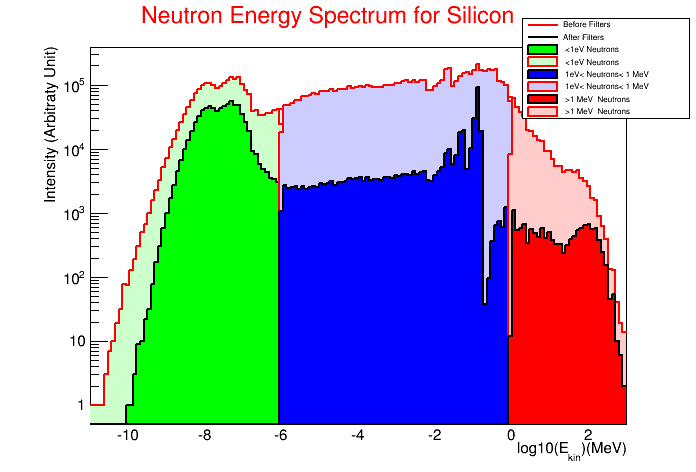}
        \caption{Neutron spectrum before and after the filter for 3 blocks of 10 cm.}
         \label{fig:sie}
    \end{subfigure}
      \hspace{2.5cm}
 \begin{subfigure}[t]{0.35\textwidth}
        \includegraphics[width=6.8cm,height=5.8cm]{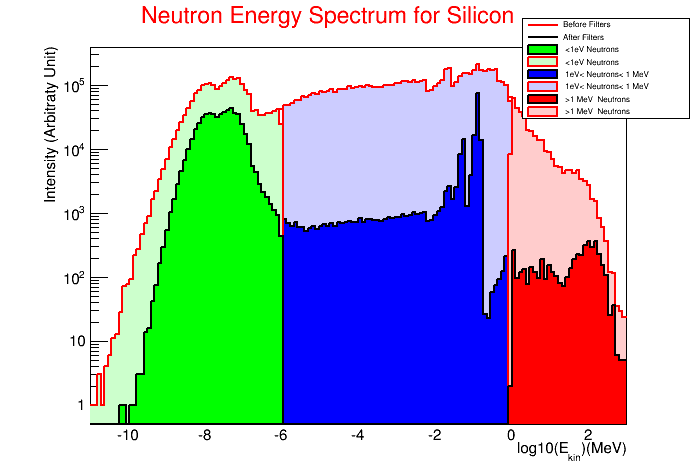}
        \caption{Neutron spectrum before and after the filter for 3 blocks of 15 cm.}
         \label{fig:sif}
    \end{subfigure}
     \caption{Neutrons energy spectra before and after the silicon filters for different geometries. }
       \label{fig:sispectrum}
\end{figure}

\begin{figure}
\centering
      \begin{subfigure}[t]{0.35\textwidth}
         \centering
               \includegraphics[width=6.8cm,height=5.8cm]{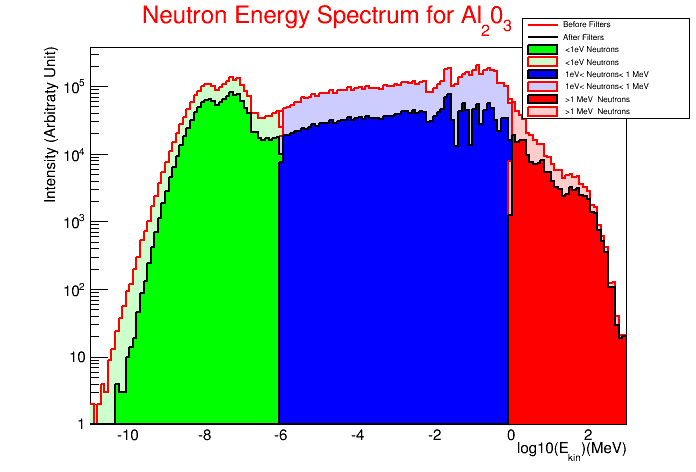}
        \caption{Neutron spectrum before and after the filter for 1 block of 2 cm. }
        \label{fig:sapa}
    \end{subfigure}
    \hspace{2.5cm}
       \begin{subfigure}[t]{0.35\textwidth}
        \includegraphics[width=6.8cm,height=5.8cm]{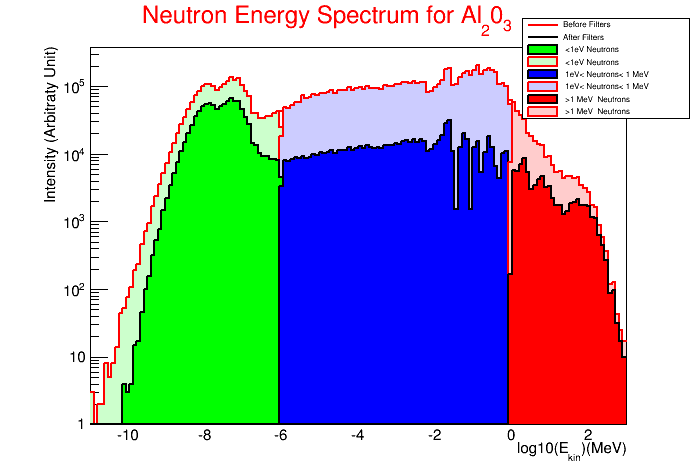}
        \caption{Neutron spectrum before and after the filter for 1 block of 5 cm.}
         \label{fig:sapb}
       \end{subfigure}
         \begin{subfigure}[t]{0.35\textwidth}
        \includegraphics[width=6.8cm,height=5.8cm]{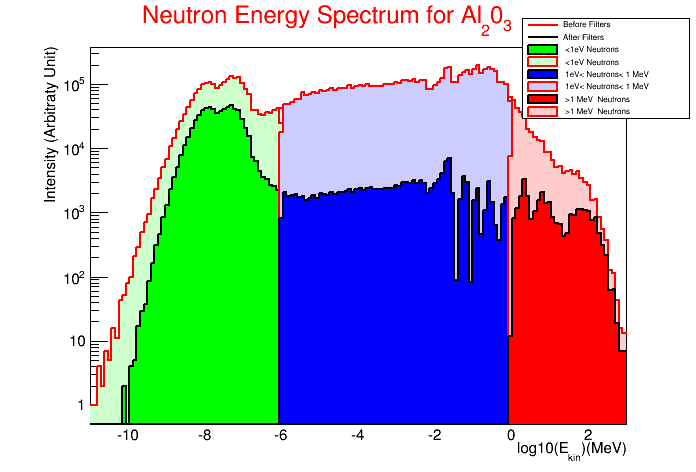}
        \caption{Neutron spectrum before and after the filter for 1 block of 10 cm.}
        \label{fig:sapc}
        \end{subfigure}
        \hspace{2.5cm}
       \begin{subfigure}[t]{0.35\textwidth}
        \includegraphics[width=6.8cm,height=5.8cm]{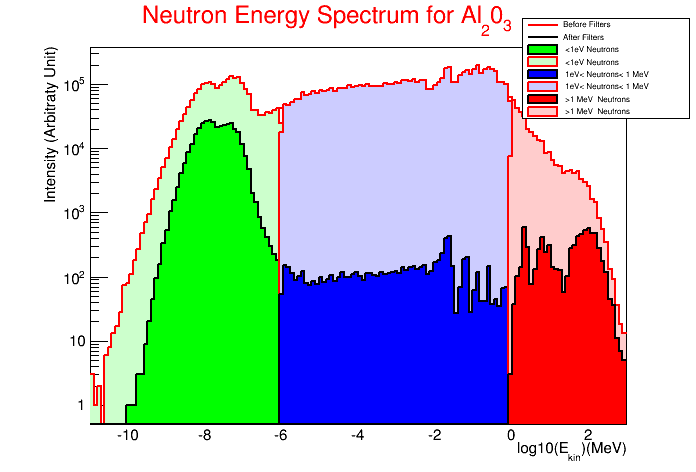}
        \caption{Neutron spectrum before and after the filter for 2 blocks of 10 cm.}
         \label{fig:sapd}
    \end{subfigure}
    \begin{subfigure}[t]{0.35\textwidth}
        \includegraphics[width=6.8cm,height=5.8cm]{EkinVaccum_neut_log_full_2_10_Al203.png}
        \caption{Neutron spectrum before and after the filter for 3 blocks of 10 cm.}
         \label{fig:sapd}
    \end{subfigure}
      \hspace{2.5cm}
 \begin{subfigure}[t]{0.35\textwidth}
        \includegraphics[width=6.8cm,height=5.8cm]{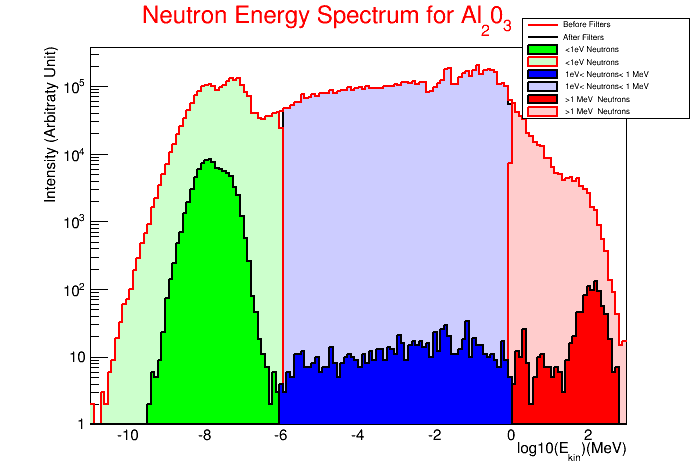}
        \caption{Neutron spectrum before and after the filter for 3 blocks of 15 cm.}
         \label{fig:sapd}
    \end{subfigure}
     \caption{Neutrons energy spectra before and after the sapphire filters for different geometries. }
       \label{fig:sapspectrum}
\end{figure}
\section{Simulation of  single crystal filter}
\begin{figure}
    \centering
    \begin{subfigure}[b]{0.35\textwidth}
        \includegraphics[width=6.8cm,height=5.6cm]{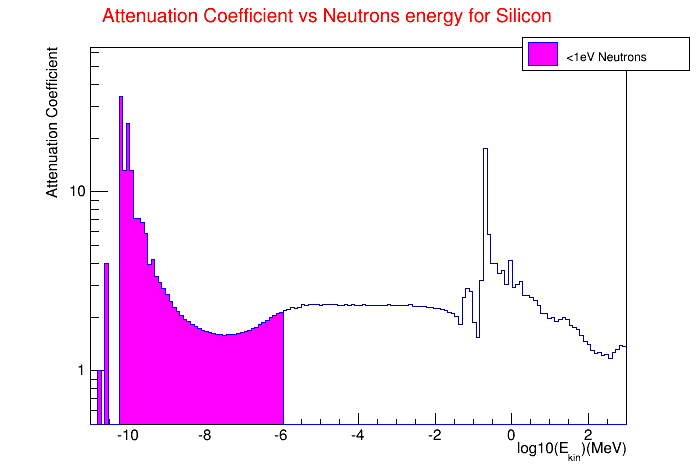}
        \caption{Attenuation Coefficient versus neutron energy for a filter composed by 1 block of 2 cm. }
        \label{fig:sapa}
    \end{subfigure}
       \hspace{2.5cm}
       \begin{subfigure}[b]{0.35\textwidth}
        \includegraphics[width=6.8cm,height=5.6cm]{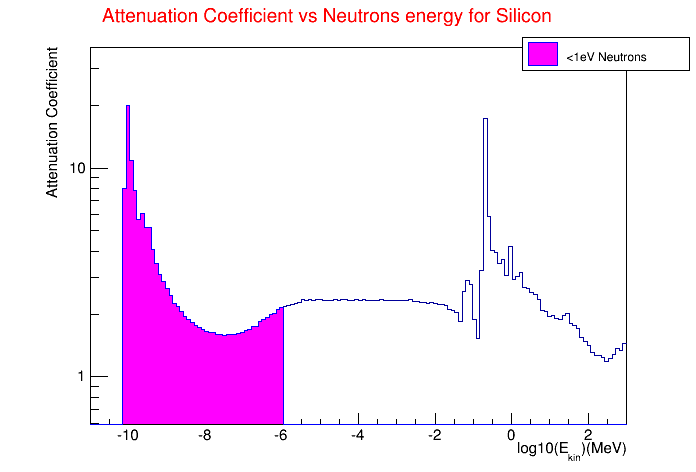}
        \caption{Attenuation Coefficient versus neutron energy for a filter composed by 1 block of 5 cm.}
         \label{fig:sapb}
       \end{subfigure}
         \begin{subfigure}[b]{0.35\textwidth}
        \includegraphics[width=6.8cm,height=5.6cm]{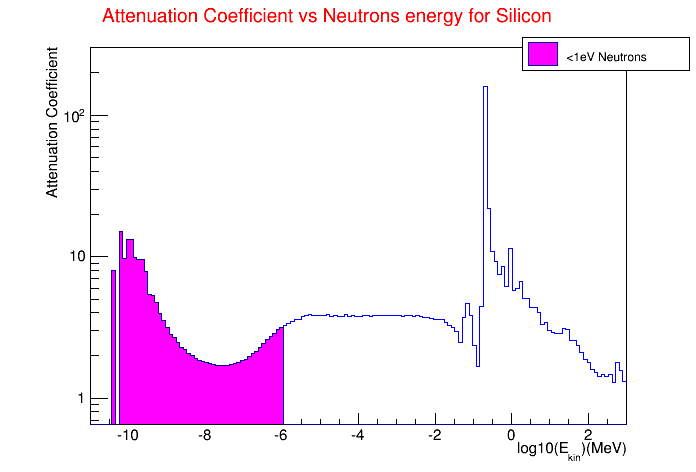}
        \caption{Attenuation Coefficient versus neutron energy for a filter composed by 1 block of 10 cm.}
        \label{fig:sapc}
        \end{subfigure}
           \hspace{2.5cm}
       \begin{subfigure}[b]{0.35\textwidth}
        \includegraphics[width=6.8cm,height=5.6cm]{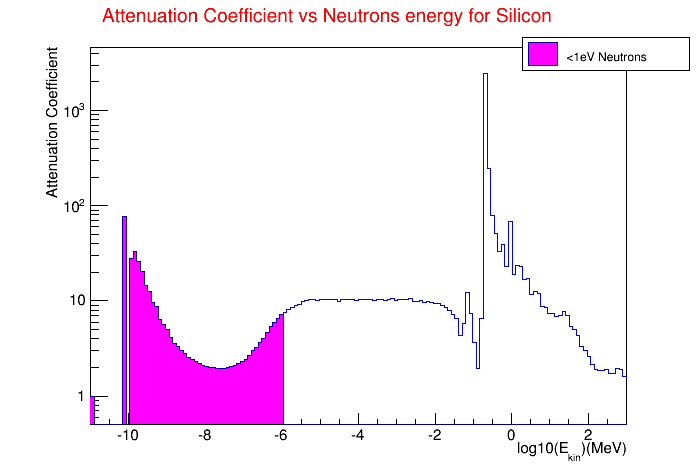}
        \caption{Attenuation Coefficient versus neutron energy for a filter composed by 2 blocks of 10 cm.}
                 \label{fig:sapd}
    \end{subfigure}
  \begin{subfigure}[b]{0.35\textwidth}
        \includegraphics[width=6.8cm,height=5.6cm]{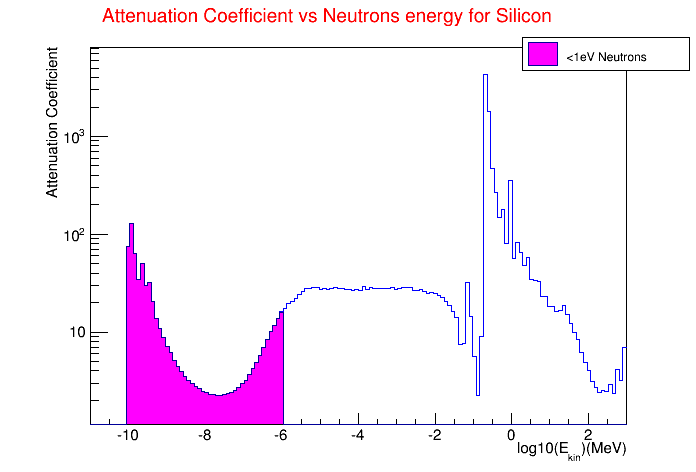}
        \caption{Attenuation Coefficient versus neutron energy for a filter composed by 3 blocks of 10 cm.}
                 \label{fig:sapd}
    \end{subfigure}
       \hspace{2.5cm}
  \begin{subfigure}[b]{0.35\textwidth}
        \includegraphics[width=6.8cm,height=5.6cm]{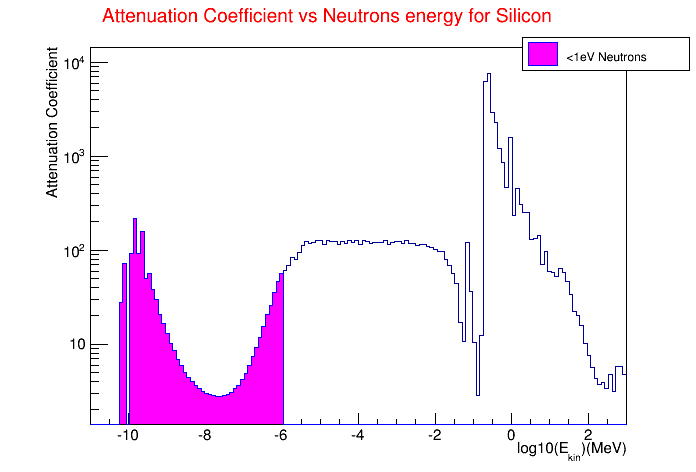}
        \caption{Attenuation Coefficient versus neutron energy for a filter composed by 3 blocks of 15 cm.}
                 \label{fig:sapd}
    \end{subfigure}
     \caption{Attenuation Coefficient versus neutron energy for silicon filters for different geometries. }
       \label{fig:siliconatt}
\end{figure}

\begin{figure}
    \centering
    \begin{subfigure}[b]{0.35\textwidth}
        \includegraphics[width=6.8cm,height=5.6cm]{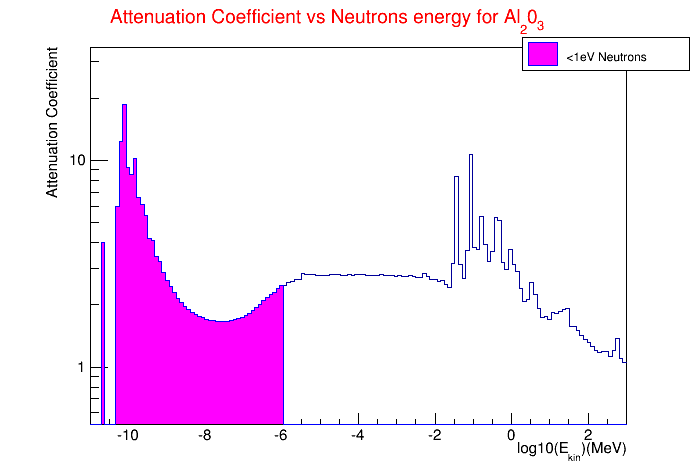}
        \caption{Attenuation Coefficient versus neutron energy for a filter composed by 1 block of 2 cm.}
        \label{fig:sapa}
    \end{subfigure}
       \hspace{2.5cm}
       \begin{subfigure}[b]{0.35\textwidth}
        \includegraphics[width=6.8cm,height=5.6cm]{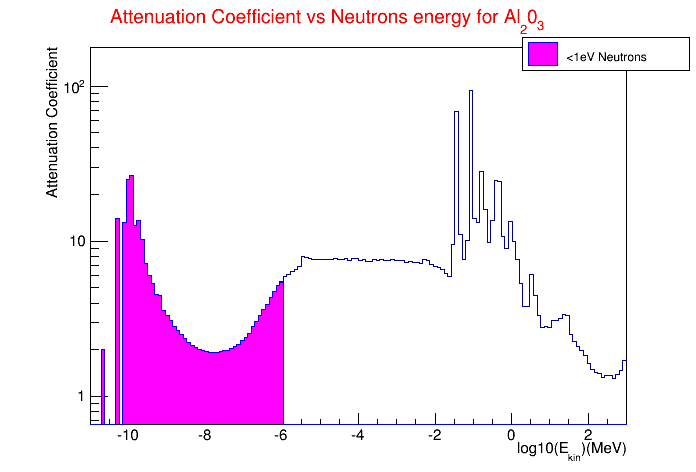}
        \caption{Attenuation Coefficient versus neutron energy for a filter composed by 1 block of 5 cm.}
         \label{fig:sapb}
       \end{subfigure}
         \begin{subfigure}[b]{0.35\textwidth}
        \includegraphics[width=6.8cm,height=5.6cm]{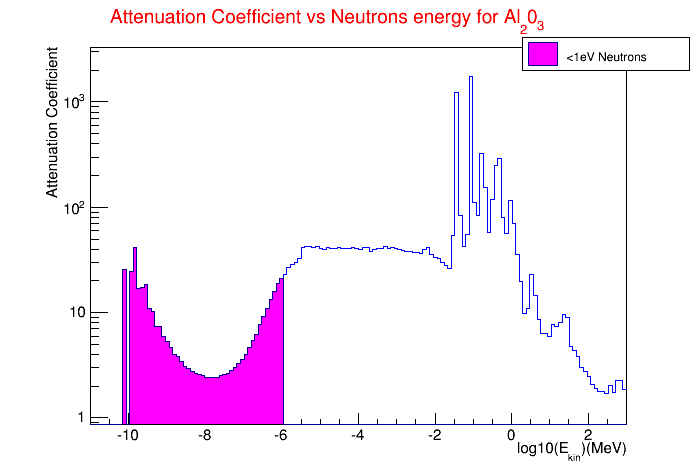}
        \caption{Attenuation Coefficient versus neutron energy for a filter composed by 1 block of 10 cm.}
        \label{fig:sapc}
        \end{subfigure}
           \hspace{2.5cm}
       \begin{subfigure}[b]{0.35\textwidth}
        \includegraphics[width=6.8cm,height=5.6cm]{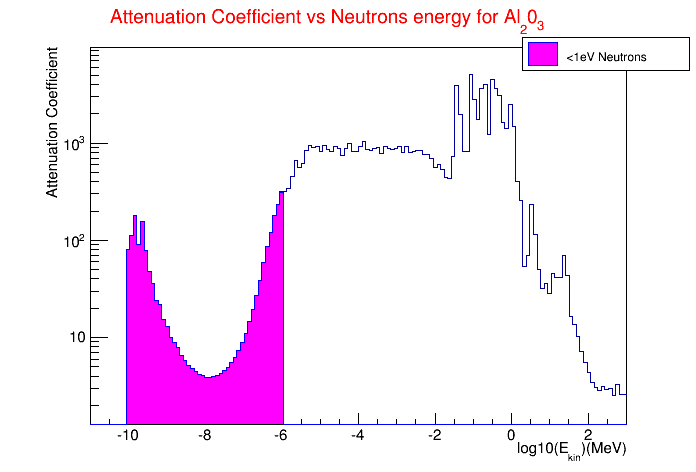}
        \caption{Attenuation Coefficient versus neutron energy for a filter composed by 2 blocks of 10 cm.}
                 \label{fig:sapd}
    \end{subfigure}
  \begin{subfigure}[b]{0.35\textwidth}
        \includegraphics[width=6.8cm,height=5.6cm]{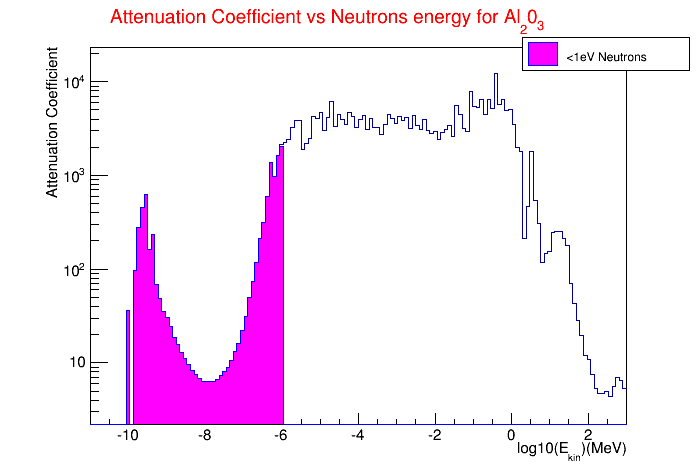}
        \caption{Attenuation Coefficient versus neutron energy for a filter composed by 3 blocks of 10 cm.}
                 \label{fig:sapd}
    \end{subfigure}
       \hspace{2.5cm}
  \begin{subfigure}[b]{0.35\textwidth}
        \includegraphics[width=6.8cm,height=5.6cm]{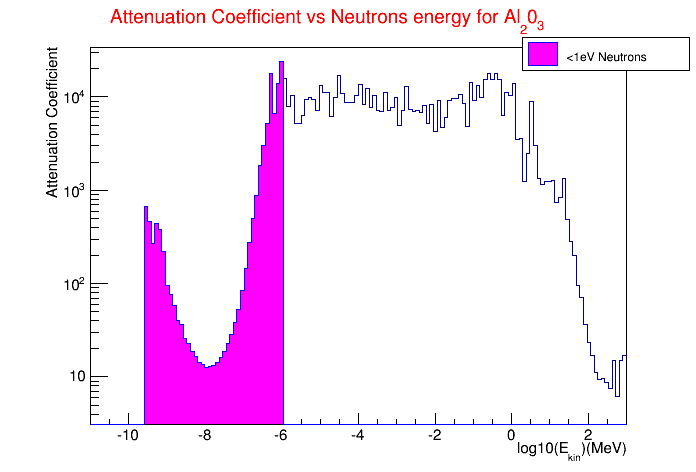}
        \caption{Attenuation Coefficient versus neutron energy for a filter composed by 3 blocks of 15 cm.}
                 \label{fig:sapd}
    \end{subfigure}
     \caption{Attenuation Coefficient versus neutron energy for \sap~filters for different geometries. }
       \label{fig:sappatt}
\end{figure}

\begin{figure}
    \centering
        \begin{subfigure}[b]{0.5\textwidth}
        \includegraphics[width=8.5cm,height=6.7cm]{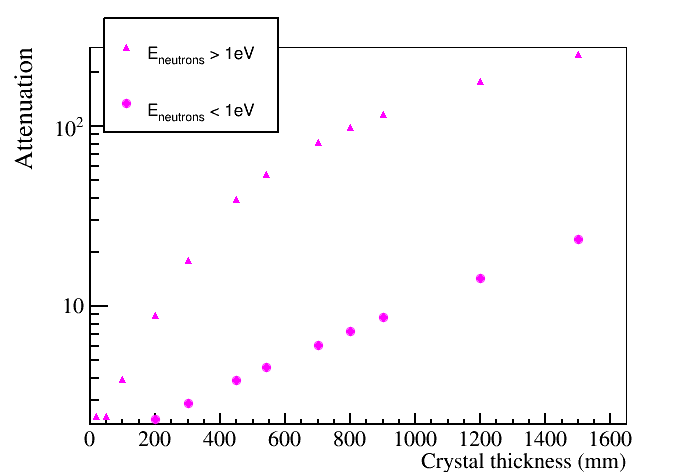}
        \caption{Attenuation Coefficient vs crystal thickness (mm).}
         \label{fig:sapb}
       \end{subfigure}
         \begin{subfigure}[b]{0.5\textwidth}
        \includegraphics[width=8.5cm,height=6.7cm]{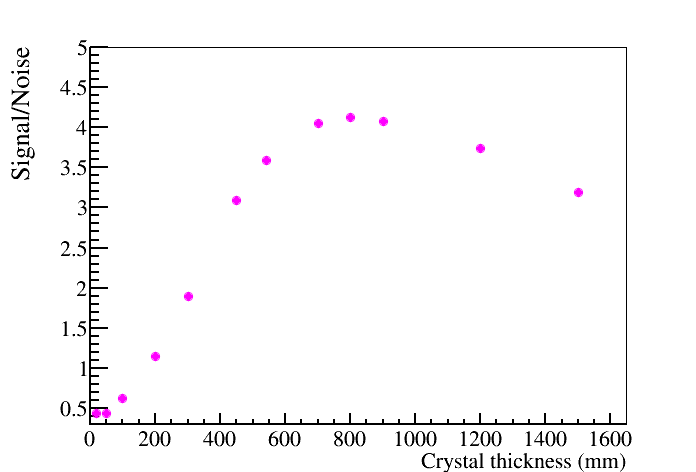}
        \caption{ Signal to noise ratio vs crystal thickness (mm).}
        \label{fig:sapc}
        \end{subfigure}
     \caption{Attenuation Coefficient and Signal to noise ratio for different thicknesses (mm) of Si filter.s}
       \label{fig:signaltonoisesi}
\end{figure}

\begin{figure}
    \centering
        \begin{subfigure}[b]{0.5\textwidth}
        \includegraphics[width=8.5cm,height=6.7cm]{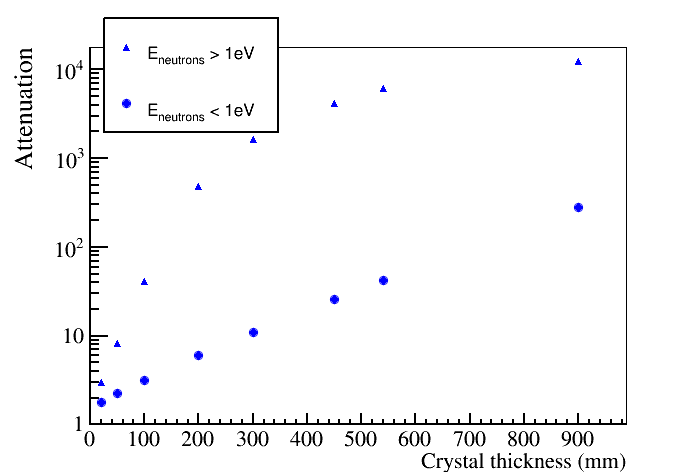}
        \caption{Attenuation Coefficient vs crystal thickness (mm).}
         \label{fig:sapb}
       \end{subfigure}
         \begin{subfigure}[b]{0.5\textwidth}
        \includegraphics[width=8.5cm,height=6.7cm]{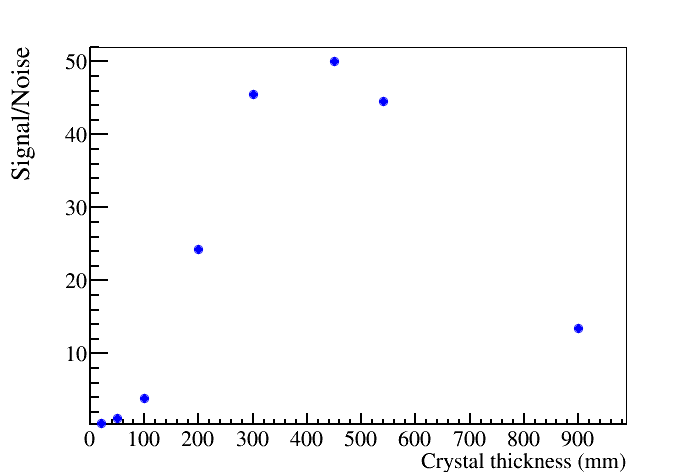}
        \caption{ Signal to noise ratio vs crystal thickness (mm).}
        \label{fig:sapc}
        \end{subfigure}
     \caption{Attenuation Coefficient and Signal to noise ratio for different thicknesses (mm) of \sap~filters.}
       \label{fig:signaltonoise}
\end{figure}

To find the optimum geometrical layout and the best material choice for ESS we simulated a straight beam line using different single crystal block configurations.
A neutron beam with the spectrum shown in Fig.~\ref{fig:modspectrum}  is used as a source and the filter is represented by different shielding blocks as shown in Fig.~\ref{fig:geant4sim}.
These blocks are surrounded by a thin layer (2cm) of B$_{4}$C to absorb the thermal neutrons and to avoid activation of the material.
At the end of the beam line is a detector to record the spectrum that has crossed the filters.
We tested several different configurations of the filters (1 block, 2 blocks, 3 blocks) with different thickness of Si and \sap. The Si is cooled to 50K while the \sap~is at room temperature.
In table \ref{tab:sil} and \ref{tab:sap}  is a summary of the obtained results, while Fig.~\ref{fig:sispectrum} and  \ref{fig:sapspectrum}  show the comparison of the ESS spectrum before and after the filter, respectively for Si and \sap. The thickness necessary to have a good attenuation of the high energy component is larger for Silicon than compared to sapphire, but even in the case of the sapphire 
the best configuration corresponds to 3 blocks of 15 cm, which is quite large if compared to what is currently used at neutron facilities~\cite{sapphirefiltersel1}. 
This is due to the fact that the ESS TDR spectrum has a large energy range 
that is difficult to attenuate with one single crystal block.
Nevertheless, for some configurations like the 3 blocks of 15 cm of \sap~the S/N ratio improved significantly, see Fig.~\ref{fig:signaltonoise}.\\
\indent As stated before, the intrinsic property of the filters is acting in a different way for different neutron energies and this feature can be seen clearly in Fig.~\ref{fig:siliconatt} and  Fig.~\ref{fig:sappatt}
where the attenuation coefficient versus neutron energy is shown for Si and \sap. A clear drop in the attenuation coefficient is seen in the energy range of interest of the ESS neutron instruments followed by a rise at the energy range of the background neutrons. The sapphire filter seems to work better than the Silicon filter, but it is interesting to note for the last one its behaviour in the region of the neutron resonance
around 100 keV where the interference between resonance and potential scattering create a window in the cross section followed by a peak (Fig.~\ref{fig:sid}~\ref{fig:sie}~\ref{fig:sif}).\\
\indent Fig.~\ref{fig:signaltonoisesi} and Fig.~\ref{fig:signaltonoise} show the attenuation coefficient for ``Signal" neutrons ( $<$ 1~eV) and ``Background" ( neutrons with energy $>$1~eV) and the S/N ratio for different thickness for Silicon and \sap~filters. As can be seen for the first, the best ratio is the configuration of 3 blocks of 27 cm while for the second one it is 3 blocks of 15 cm.
\section{Conclusions }
We showed the capabilities of Geant4 with the additional package NCrystal to simulate single crystal filters.
We studied the optimum filter for an ESS beam line using the ESS TDR moderator spectrum.
The results of our studies show that single crystal filters can improve the S/N ratio by 10 to 100 times at a reduced cost of absolute flux. 
Combinations of \sap~with a Silicon layer could be advantageous and need additional studies.
  A filter is a viable component on both curved and straight instruments but optimization of the filter and beam line geometry must be done separately for each beam line taking into account the specific requirements of the instrument in question.

\end{document}